\documentclass[11pt,final]{article}

\usepackage{textgreek}

\makeatletter
\def\set@curr@file#1{%
    \begingroup
    \escapechar\m@ne
    \xdef\@curr@file{\expandafter\string\csname #1\endcsname}%
    \endgroup
}
\makeatother

\usepackage{multirow}
\usepackage{amsmath}
\usepackage{amsthm}
\usepackage[dvipsnames,table,xcdraw]{xcolor}
\usepackage[english]{babel} 
\usepackage[toc,page]{appendix}
\usepackage{amssymb}
\usepackage[centercolon]{mathtools}
\usepackage[textsize=footnotesize]{todonotes}
\usepackage[utf8]{inputenc}
\usepackage[T1]{fontenc}
\usepackage{mathptmx} 
\usepackage[scaled=0.82]{beramono}
\usepackage{tabularx}
\usepackage{authblk}
\usepackage{rotating}
\usepackage{graphicx}
\usepackage{here}
\usepackage{booktabs}
\usepackage{pdfpages}
\usepackage[onehalfspacing]{setspace}
\usepackage{cases}
\usepackage{lmodern}
\usepackage[normalem]{ulem}
\usepackage{bm}
\usepackage{parskip}

\makeatletter
\def\tagform@#1{\maketag@@@{[\ignorespaces#1\unskip\@@italiccorr]}}
\makeatother

\usepackage{subcaption}

\usepackage[paperwidth=8.5in, paperheight=11in, margin=1in]{geometry}
\usepackage{cite}

\usepackage{ifdraft}

\newcommand{\stkout}[1]{%
\ifmmode
    \text{\sout{\ensuremath{#1}}}%
\else
    \sout{#1}%
\fi
}
\newcommand{\R}[1]{\ifoptionfinal{}{\marginpar{{\color{red}#1}}}}
\newcommand{\B}[1]{\ifoptionfinal{#1}{{\color{blue}#1}}}
\newcommand{\delete}[1]{\ifoptionfinal{}{{\color{red}\sout{#1}}}}
\newcommand{\abs}[1]{\ensuremath{\left\lvert#1\right\rvert}}

\date{}

\usepackage{hyperref}
\usepackage{cleveref}

\begin{document}

\title{Enhanced and Robust Contrast in CEST MRI: Saturation Pulse Shape Design via Optimal Control}

\author{Clemens Stilianu$^1$, Christina Graf$^1$, Markus Huemer$^1$, Clemens Diwoky$^2$, Martin Soellradl$^{1,3}$, Armin Rund$^{4}$, Moritz Zaiss$^{5,6}$, and Rudolf Stollberger$^{1,7}$}

\maketitle

\begin{center}
\small{$^1$Institute of Biomedical Imaging, Graz University of Technology, Graz, Austria\\$^2$Institute of Molecular Biosciences, University of Graz, Graz, Austria
\\$^3$Department of Radiology and Radiological Sciences, 
Monash University, Melbourne, Australia\\$^4$Institute for Mathematics and Scientific Computing, University of Graz, Graz, Austria\\$^5$Institute of Neuroradiology, Friedrich-Alexander-Universität Erlangen-Nürnberg (FAU), University Hospital Erlangen, Erlangen, Germany\\$^6$High-Field Magnetic Resonance Center, Max-Planck Institute for Biological Cybernetics, Tübingen, Germany\\$^7$BioTechMed Graz, Graz, Austria}
\end{center}

\hfill 
\hfill
 
\textbf{Running head:} Enhanced and Robust Contrast in CEST MRI: Saturation Pulse Shape Design via Optimal Control\\

\textbf{Correspondence to:} Rudolf Stollberger - rudolf.stollberger@tugraz.at\\

\textbf{Funding:} This research was funded in whole, or in part, by the Austrian Science Fund (FWF) I4870. For the purpose of open access, the author has applied a CC BY public copyright license to any Author Accepted Manuscript version arising from this submission.\\

\textbf{Word Count} (approximately) 243 (Abstract) \delete{<5300}\B{<5500} (Body) \\

Submitted to \textit{Magnetic Resonance in Medicine} as a Research Article.\\
Part of this work has been presented at the ISMRM Annual Conferences in 2022 and 2023.

\clearpage

\section*{Abstract}

Purpose: To employ optimal control for the numerical design of CEST saturation pulses to maximize contrast and stability against $B_0$ inhomogeneities. \\

Theory and Methods:
We applied an optimal control framework for the design pulse shapes for CEST saturation pulse trains. The cost functional minimized both the pulse energy and the discrepancy between the corresponding CEST spectrum and the target spectrum based on a continuous RF pulse. The optimization is subject to hardware limitations. In measurements on a 7 T preclinical scanner, the optimal control pulses were compared to continuous-wave and Gaussian saturation methods. We conducted a comparison of the optimal control pulses were compared to with Gaussian, block pulse trains, and adiabatic spin-lock pulses.  \\

Results:
The optimal control pulse train demonstrated saturation levels comparable to continuous-wave saturation and surpassed Gaussian saturation by up to 50 \% in phantom measurements. In phantom measurements at 3 T the optimized pulses not only showcased the highest CEST contrast, but also the highest stability against field inhomogeneities. In contrast, block pulse saturation resulted in severe artifacts. Dynamic Bloch-McConnell simulations were employed to identify the source of these artifacts, and underscore the $B_0$ robustness of the optimized pulses. \\

Conclusion:
In this work, it was shown that a substantial improvement in pulsed saturation CEST imaging can be achieved by using Optimal Control design principles. It is possible to overcome the sensitivity of saturation to B0 inhomogeneities while achieving CEST contrast close to continuous wave saturation.

\subsection*{Keywords} Optimal Control, Pulsed CEST, Pulse Design, B0 Robustness, Contrast Mechanism
\clearpage

\section{Introduction}
Chemical Exchange Saturation Transfer (CEST) is an emerging MRI technique that allows for detection of molecules via protons that are in chemical exchange with the surrounding water protons. This exchange transfers saturation from molecules to water, which can be detected indirectly by standard MRI methods. This enables fast and high-resolution images that contain spectroscopic information in every voxel. In CEST imaging, the signal obtained from a mobile proton pool gets amplified, which allows for the detection of low concentration molecules \cite{ward2000new, wu2016overview}. Contrast of relevance for clinical applications can be derived from groups such as amide, amine, and hydroxyl in mobile protons and peptides, as well as from glycosaminoglycans, glutamate, and shift in creatine and pH levels \cite{yan2015assessing, zhou2003using, harris2015ph, zhou2006chemical, cai2012magnetic, kogan2014method, wang2016magnetic}. Early uses of CEST MRI in medical environments have shown promise for evaluating various medical conditions, encompassing strokes, cancer, osteoarthritis, muscle function, lymphedema, multiple sclerosis, Alzheimer's disease and other neurological pathologies. \cite{jones2018clinical, goldenberg2019assessments} \\

A typical CEST experiment consists of the application of an off-resonant saturation pulses with a duration of a few seconds, followed by an imaging sequence. This is repeated for various off-resonant frequencies to accumulate a CEST- or Z-spectrum. The quality of the Z-spectrum is not only crucial for the acquisition of accurate spectroscopic information, but also for the reliability and resilience of the resulting image data. One of the most important parameters influencing the characteristics of the Z-spectrum is the saturation strategy. A common approach in preclinical research is continuous wave (CW) saturation, which result in a spectrum with high exchange weighting. However, on clinical systems, CW pulses with a duration of several seconds cannot be applied due to hardware limitations. Consequently, pulsed CEST saturation is the standard on clinical scanners. \cite{smith2006pulsed,schmitt2011optimization,CWbetter1,CWbetter2}. \\

The optimization of RF pulse parameters within a pulsed CEST experiment to increase CEST contrast was explored in the past \cite{schmitt2011optimization,xiao2015fast,kim2019optimization}. Nevertheless, a large number of pulsed CEST applications still rely on predefined saturation pulse shapes, such as Gaussian or a slight modification of Gaussian line shapes as well as Block Pulse Trains (BPT) \cite{zu2011optimizing}. The research conducted by Yoshimaru et al. indicates that the shape of the saturation pulse is a critical parameter to consider when striving for optimal CEST spectrum quality \cite{yoshimaru2016design}. With the off-resonance Spin Lock (SL) method, an improved pulsed saturation technique was introduced \cite{SLoverCEST}. To reduce the sensitivities to field inhomogeneities, the tip-down and tip-up pulses were substituted with optimized adiabatic pulses for certain $B_0$ field strengths \cite{herz2019cest,herz2019t1rho}. \\ 

RF pulse design by Optimal Control (OC) has been used in the past to design RF pulses within a constrained parameter space. By accounting for physiological and technical restrictions, this approach ensures the feasibility of RF pulses on clinical scanners, while minimizing certain parameters such as pulse duration, specific absorption rate or $B_1$/$B_0$ sensitivity \cite{conolly1986optimal,graf2022advanced,rund2018simultaneous,aigner2020time,rund2017magnetic,rund2018optimal}. \\

In this work, we design the RF pulse shape of a CEST saturation pulse train by OC. The design criteria were chosen to attain high quality saturation, including high exchange weighting and stability against field inhomogeneities. We compare the OC pulses with CW saturation on a 7T preclinical scanner in phantom measurements. We further compare our pulses with Gaussian, BPT and adiabatic SL pulses on 3T in phantom measurements. Moreover, we perform Bloch-McConnell simulations of the magnetization vector dynamics with OC and BPT to understand their behavior under the influence of $B_0$ inhomogeneities. \\

\pagebreak

\section{Methods}

\subsection{RF pulse optimization and Bloch-McConnell simulation}

The RF magnitude, $B_1(t)$, of a pulse train with a saturation time of $T_{sat}$, duration of a single pulse $t_d$ and pause between pulses $t_p$ is optimized by minimizing the objective function,

\begin{flalign}
    \underset{B_1(t)} {\min}\, & J(B_1(t), M_z(\omega,t),\tilde{M}_z(\omega,t)) =  & \nonumber \\ & \frac{\alpha}{2}\int\limits_{t=0}^{T_{sat}}B_1(t)^{2} dt
    + \frac{\sigma_1}{p}\sum_{\omega} \abs{\frac{M_z(\omega,T_{sat})-{M}_{zdes}(\omega)}{\epsilon}}^p 
    + \frac{\sigma_2}{p}\sum_{\omega} \abs{ \frac{\tilde M_z(\omega,T_{sat}) - \tilde{M}_{zdes}(\omega)}{\epsilon}} ^p ,
\label{cost_functional}
\end{flalign}

\begin{align}
\text{s.t.}
\begin{cases}
    0 \leq B_1(t) \leq B_{1max} \\
    \frac{dM(\omega,t)}{dt} = A \cdot M(\omega,t) + b, \ \frac{d\Tilde{M}(\omega,t)}{dt} = \Tilde{A} \cdot \Tilde{M}(\omega,t) + b, \ \forall \omega \in \Omega, \ \forall t \in (0,T_{sat}) \\
\end{cases}
\end{align}

The first term of the cost functional minimizes the pulse energy with regularization parameter $\alpha>0$. While $M_z(\omega,T_{sat}), \ {M}_{zdes}(\omega)$ denote the spectra at the end of the pulse with two pools, $\tilde M_z(\omega,T_{sat}), \ \tilde{M}_{zdes}(\omega)$ represent the spectra with solely water pool. $A$ is the Bloch-McConnell system matrix depending on RF pulse $B_1(t)$. $M(\omega,t)$ and $\tilde{M}(\omega,t)$ are the solutions of the Bloch-McConnell equations for the off-resonance $\omega$ in the z-spectrum $\Omega$. The simulation is discretized in time $t$ with a step size of $\Delta t = 100 \ \mu s$. Temporal discretization was chosen based on \cite{Christina1}. The second and third terms minimize the difference between the CEST spectrum, ${M}_{z}$, at the end of the RF pulse and a desired target CW CEST spectrum, ${M}_{zdes}$. $\sigma_{1,2}>0$ are weighting parameters. The metric for the differences is conducted via the $L^p$-norm, $p \geq 2$ even. $\epsilon$ specifies the allowed deviation from the target magnetization. We loop over $p$ starting with $p=2$ increasing the value successively as proposed by \cite{rund2017magnetic}, since a lower value of $p$ has been found to confer advantages for globalization. Optimization was constrained to the points during $t_d$. The pause $t_p$ was set to $B_1(t)$ = 0. \\

The numerical optimization itself is based on a trust-region semi-smooth quasi-Newton method \cite{rund2017magnetic}. Exact discrete derivatives are supplied by adjoint calculus \cite{troltzsch2005optimale}. First, the adjoint equations to the Bloch-McConnell equations are introduced as 

\begin{align}
\begin{cases}
    \frac{-dq(\omega,t)}{dt}-A^Tq(\omega,t)=0 \ \text{in} \ (0,T_{sat}) \\
    q(\omega,T_{sat})=(0,0,\frac{\sigma_1}{\epsilon} (\frac{M_z(T_{sat},\omega)-M_{zdes}(\omega)}{\epsilon})^{(p-1)}).
\end{cases}
\end{align}

Here, $q$ is the adjoint state corresponding to $M$. Analogously the other adjoint state $\Tilde{q}$ can be derived with $\Tilde{M}$, in which only the water pool is active. Subsequently, the reduced gradient $g$ of the cost functional $J$ can be identified as 

\begin{align}
    g(B_1) = \alpha B_1 + \int\limits_{\Omega}^{} q^TA_{B_1} M d\omega + \int\limits_{\Omega}^{} \Tilde{q}^T A_{B_1} \Tilde{M} d\omega
\end{align}

whereby 

\begin{align} 
    A_{B_1} = \begin{pmatrix}
    0 & 0 & 0\\
    0 & 0 & \gamma\\
    0 & -\gamma & 0
    \end{pmatrix} 
\end{align}

The CEST-spectra are simulated using two-pool Bloch-McConnell equations solved numerically by symmetric operator splitting \cite{Christina1}. \\

For simulation and optimization, the water pool was set to relaxation times $T_{1w} =$ 1000 ms and $T_{2w}=$ 80 ms. The solute pool was simulated to resemble a creatine phantom with relaxation times $T_{1s} =$ 1000 ms and $T_{2s} =$ 160 ms and a fraction rate of $f_s =$ 0.002 \cite{zaiss2011quantitative}. The exchange between the solute pool and the water pool was set to $k_{sw} =$ 210 s$^{-1}$. The exchange was estimated for creatine at room temperature and pH 7.4 with the formula suggested in \cite{goerke2014characterization}. The $B_0$ field was 3~T. The spectrum was optimized for $\Omega \in [-6, 6]$ ppm, discretized in $\Delta \omega =$ 0.02 ppm steps with a total number of spatial points of $N_\omega = 601$. ${M}_{zdes}$ was simulated with the same parameters for a $T_{sat} =$ 1 s CW saturation pulse with an RF amplitude of $B_1 =$ 1 \textmu T. The optimized pulse was constrained to a saturation time of $T_{sat} =$ 1 s and had a duty cycle of 90 \%, a pulse duration time of $t_{d} =$ 100 ms, and $t_{p} =$ 12.5 ms pauses between each of the nine pulses (Pulses optimized with different DC, $T_{sat}$, $t_d$ and $t_p$ can be found in the Supporting Information Figures S5-8). Optimization was constrained to the points during $t_d$. The pause $t_p$ was set to $B_1(t)$ = 0. The RF amplitude was constrained to $0 \leq B_1 \leq 3$ \textmu T. The regularization was $\alpha = 100 \ \sigma_1$ and $\sigma_1 = \sigma_2$. We loop over $p$ starting with $p=2$ increasing the value successively as proposed by \cite{rund2017magnetic}, since a lower value of $p$ has been found to confer advantages for globalization. The optimization was initialized with a BPT and converged with $p$ = 4. \\

\subsection{Evaluation of OC pulse in simulations}
We evaluated the impact of various parameters on the saturation of the OC pulse in simulations, such as $T_1$, $T_2$, exchange rate $k$, $B_0$ and $B_1$ amplitude scaling of the pulse. The performance of the OC pulse was assessed by the smoothness of the spectra, the stability of the Magnetization Transfer Ratio Asymmetry, $MTR_{asym}$, and the maximum value of $MTR_{asym}$ across a range of parameters. \\

To assess the impact of Magnetization Transfer (MT) on the performance of the OC pulses in comparison to Gaussian and BPT saturation, simulations were carried out using a three-pool model at 3 T. The MT pool was based on the WM parameters described in \cite{van2018magnetization}.

\subsection{Experimental setup}
Performance of OC saturation was compared to a CW pulse and Gaussian pulse train. These pulses were implemented on a 7 T preclinical scanner system (Bruker BioSpec 70/20 USR, Ettlingen, Germany). The CW pulse was applied with $T_{sat}=$ 1 s and $B_{1,RMS}=$ 1 \textmu T. The $B_1$ level was selected based on simulations showing that both CW and Gaussian pulses approximately yield the maximum creatine CEST contrast for the given saturation time. The measurement spectra were compared to two-pool Bloch McConnell simulations. The parameters were the same as for the optimization, with the exception of the relaxation times, which were adjusted to the measured $T_1$ and $T_2$ times of the phantom. CEST-spectra were measured on a 44 mM creatine monohydrate phantom in a 15-ml falcon tube. Relaxation times were reduced with Manganese(II)chloride and agarose (1 \%) to quasi-in vivo conditions \cite{rooney2007magnetic} ($T_{1}=$ (1730 $\pm$ 16) ms, $T_{2}=$ (74 $\pm$ 3 ms)), with phosphate buffer at constant pH 7.0. The spectra were measured at room temperature for $\omega\in[-3,3]$ ppm with increments of 0.05 ppm. Images were acquired with a single-shot RARE-sequence with a TR of 7 s and a $T_E$ of 40 ms with partial Fourier reconstruction. The temperature of the phantom was between 21 and 22 $^\circ$C. \\

Phantom measurements were performed on a Siemens Vida 3 T clinical scanner system (Siemens Healthineers, Erlangen, Germany) with a 20 channel head coil. OC, SL, Gauss and BPT were implemented in Pulseq-CEST (Pulseq release 1.3.1) \cite{Pulseq,Pulseq-CEST}. The saturation time was set to $T_{sat}=$ 1 s, with a Duty Cycle ($DC$) of 90 \%. The $B_{1,RMS}$ was set to 1 \textmu T, this means that all pulse trains had the same energy and therefore the same specific absorption rate (SAR). The OC and Gauss train were implemented using a constant phase. The phase of the BPT was adjusted after each pause by $\omega_n T_{pulse}  B_0 \gamma$, with the current offset $\omega_n$, to ensure a smooth spectrum of the BPT. The adiabatic SL pulse was provided by the Pulseq-CEST library. This pulse was optimized for 3 T as described by Kai Herz et al. in \cite{herz2019t1rho}. The adiabatic tilting pulses were factored into the calculation of the pulse block duration, as well as the $B_{1,RMS}$. The Gaussian saturation as it appears in Pulseq-CEST is windowed with a cosine. (For more details on the pulse trains used, see the Supporting Information Figure S1.) \\

The pulse trains were applied to a 87 mM creatine monohydrate phantom in a 500-ml glass sphere. The relaxation times were reduced with Manganese(II)chloride to $T_1=$ 680 ms, $T_2=$ 90 ms \cite{wansapura1999nmr}. The pH was stabilized at 7.0 with a phosphate buffer. Furthermore, an air bubble was placed intentionally at the top of the glass sphere to introduce $B_0$ inhomogeneities. CEST-spectra were measured for $\omega\in[-6,6]$ ppm with a step size of $d \omega =$ $\Delta \omega =$ 0.1 ppm. For readout a centric reordered 2D-GRE was used, with parameters: $FOV=128\times128$ mm$^2$, flip angle $\alpha=$ 10$^{\circ}$, $T_R=$ 12 ms, $T_E=$ 40 ms, base resolution $res=64\times64$ and slice thickness 5 mm with one slice through the center of the sphere. \\ 

In addition, to investigate the influences of the \( B_1^+ \) field on the acquired Z-spectra, a \( B_1^+ \)-map was estimated from the Bloch-Siegert (BS) shift \cite{sacolick2010b1} using a 3D-GRE readout. The sequence parameters were as follows: $T_R = $ 150 ms, $\alpha=$ 30$^{\circ}$, matrix size = $64\times 64 \times 64$, FOV = $128\times 128 \times 128$ mm$^3$, BS pulse with an off-resonance frequency of 4 kHz and a nominal BS-shift of $\phi_{BS}$ = 0.78 rad. The \( B_1^+ \)-maps were denoised using a 5 px median filter. \\\\

\subsection{Postprocessing and analysis of experimental data}

To evaluate the exchange weighting of the different pulses, $MTR_{asym}$ \cite{zhou2004quantitative,zhou2003using} was used.
A pixel-wise two pool Lorentzian fit using the MATLAB function $lsqcurvefit$ was employed to assess and correct the $B_0$ shift and to determine the peak of the CEST effect \cite{zaiss2011quantitative,jones2012vivo,dula2013amide}. To improve the visualization of structural inhomogeneity, the CEST images were denoised using TGV. \cite{shirai2014fft,bredies2010total}. \\

The geometry of the spherical glass phantom with an air bubble on top introduced different levels of \( B_1^+ \) and $B_0$ inhomogeneities into the measurement. The frequency shift in the $B_0$ map was calculated in every pixel with the water peak shift of the fitted Gauss saturation. Spectra were evaluated in different ROIs placed over the phantom to investigate different combinations of \( B_1^+ \) and $B_0$. Moreover, the robustness of the spectra and the quality of the corresponding fit were examined by inspecting the Z-spectrum of representative pixels in the inhomogeneous region. \\

\subsection{Simulation of magnetization vector dynamics of OC pulses}

To investigate and visualize the dynamic behavior of the magnetization during saturation with OC and BPT, the $x$-, $y$- and $z$-components of the magnetization were simulated over the duration of the pulse train in the water pool at -1 ppm. The simulation includes two scenarios: one in which the phase remains constant throughout the duration of the pulse, as used during the optimization process, and another in which the phase was adjusted after each pulse by $\omega_n T_{pulse} B_0 \gamma$. The latter scenario is necessary for obtaining a smooth BPT spectrum. Alterations in the phase of a pulse can be seen analogously to changes in either the pulse center frequency $\omega_n$ or the $B_0$ field (see Supporting Information for details).

\section{Results}
\subsection{Pulse optimization and simulation for different parameter ranges}

Figure \ref{parameter_check} depicts the simulation results for the OC saturation for various parameter ranges and a comparison for $MTR_{asym}$ with Gaussian and CW saturation. Apart from the varied parameters the simulation parameters from the optimization were used. In (a) saturation was simulated for $T_1 \in \{ 0.2, \ 0.5, \ 1, \ 2, \ 4 \}$ s. The spectrum and the $MTR_{asym}$ peak (beneath) remained smooth and consistent for all $T_1$ times in the simulation. In (b) $T_2$ was varied for $T_2 \in \{ 0.01, \ 0.03, \ 0.08, \ 0.15, \ 0.25 \}$ s. The OC pulse yielded a smooth spectrum when $T_2 \le $ 0.15 s. However, when $T_2$ was increased to 0.25 s, the spectrum began to exhibit wiggles. In (c) OC saturation was simulated for exchange rates $k \in \{ 50, \ 100, \ 200, \ 500, \ 1000, \ 2000 \}$ s$^{-1}$. The OC pulse produced a smooth spectrum for all values of $k$, however the $MTR_{asym}$ peak showed minor wiggles for $k$ = 2000 s$^{-1}$. In (d), the OC pulse showed a smooth spectrum for all investigated field strengths $B_0 \in \{ 1.5, \ 3, \ 5, \ 7 \}$ T. In (e), the OC pulse was scaled by a factor so that the $B_{1RMS}$ value was $B_{1RMS} \in \{ 0.3, \ 0.5, \ 0.7, \ 1, \ 1.2, \ 1.4, \ 2 \}$ \textmu T. When the scaling surpassed 200 \%, or when the $B_{1RMS}$ fell below 30 \%, the saturation began to introduce minor distortions in the spectrum. The peak value of the OC $MTR_{asym}$ demonstrated a performance comparable to that of CW saturation across all parameters examined. The maximum of $MTR_{asym}$ consistently exceeded that of Gaussian saturation at every simulated point. \\

Simulations using a three-pool model, inclusive of an MT pool, demonstrated a general decrease in $MTR_{\text{asym}}$ across CW, OC, BPT, and Gaussian saturation. Relative to the two-pool model, the MT pool led to reductions in $MTR_{\text{asym}}$ of 40.43 \% for CW, 43.97 \% for OC, 42.43 \% for BPT, and 43.93 \% for Gaussian saturation (see Supporting Information Figure S9).

\subsection{Comparison 7 T preclinical phantom measurements with simulation}
Figure \ref{fig:7T} compares results of the simulations with measurements at a field strength of 7T, for the CW, OC and Gauss pulse train. The CW pulse measurement shows a maximum $MTR_{asym}$ of 9.6 \% and the simulation result matches with a maximum of 9.5 \%. Similarly, the OC pulse measurement exhibits a maximum $MTR_{asym}$ of 9.1 \%, while the simulation result shows a maximum of 9.0 \%. The Gauss pulse train measurement demonstrates a maximum $MTR_{asym}$ of 6.7 \%, whereas the simulation result shows a maximum of 6.1 \%. The results of these measurements indicate that the exchange weighting of the CW pulse is 4.4 \% higher than that of the OC pulse and 40.3 \% higher than that of the Gauss pulse train. Analogously, the simulation results show that the exchange weighting of the CW pulse is 6.7 \% higher than that of the OC pulse and 57.4 \% higher than that of the Gauss pulse train.

\subsection{Creatine phantom measurements on a 3 T clinical scanner}
Figure \ref{fig:3T_sphere_asym} shows the results of the 3 T phantom measurements for the OC, BPT, adia SL and the Gaussian saturation pulses. (a) depicts the uncorrected CEST image at the creatine peak of the $MTR_{asym}$. For the different saturation strategies, the $B_0$ inhomogeneities had different effects on the images. Especially the BPT exhibited a high CEST peak shift and expressed wavelike behavior. The shift is particularly expressed under the air bubble and on the edge of the sphere. Across all images, the center of the sphere exhibited a relatively low frequency shift and high homogeneity. The OC, adia SL, and the Gauss saturation pulses exhibited a comparably homogeneous CEST contrast and did not express wave-like behavior in the images. After $B_0$ correction (b) and TGV denoising (c), the BPT still showed significant wave-like behavior at the largest $B_0$ offsets. The the images generated using the OC, adia SL and Gauss saturation pulses did not express these behavior. The OC pulse generated a homogeneous CEST image with high contrast, while the adia SL pulse and Gaussian pulse still displayed a slight inhomogeneous pattern. Specifically, at the bottom of the sphere, the adia SL pulse appeared to have a reduced CEST amplitude, whereas the BPT exhibited again wave-like behavior. The homogeneity increased after fitting the OC and the BPT images (d, e). In the BPT image, the fit was able to partially reduced the wavelike artifacts compared to the $MTR_{asym}$ images. The fitted adia SL image exhibited a slight degradation in homogeneity. Additionally, the fit indicated a higher exchange weighting for the Gauss pulse train, particularly at the center of the sphere. The two-pool fit showed lower noise overall. \\

The difference in exchange weighting and homogeneity among the different saturation regimes, particularly in the region of the highest $B_0$ shift, is shown by a vertical profile through the center of the sphere of the $B_0$ corrected TGV denoised $MTR_{asym}$ images (f). The wave-like behavior of the BPT is evident between pixels 10 and 30, with a slight peak in exchange weighting for the OC and adia SL and a comparable smooth Gauss saturation in this region. The exchange weighting of the Gauss pulse train is substantially lower compared to other saturation strategies, with the OC pulse exhibiting the highest exchange weighting, followed by the BPT pulse and the adia SL pulse. \\

\subsection{Influence of field inhomogeneities to the spectrum}
Figure \ref{fig:ROI} (a) illustrates the changes in $\Delta \omega$ due to $\Delta B_0$. In (b), the \( B_1^+ \) map divided by the nominal BS \( B_1 \) level, \( B_{1 BS} \), is presented along with a line profile. The profile was selected to traverse from a region with least \( \Delta \omega \) shift to one with high \( \Delta \omega \) shift, while also passing through the sphere's center to span the entire \( B_1^+ \) range. Therefore, in one half of the profile, the influence of \( B_1 \) is isolated, while in the other half, both \( B_1 \) and $B_0$ field imperfections are present. To better visualize the impact of field inhomogeneities on the saturation method, the $MTR_{asym}$ images were normalized and scaled using the maximum value of $\frac{B_1^+}{B_{1 BS}}$. The \( B_1^+ \) and the $\Delta \omega$ as well as the profile in the scaled $MTR_{asym}$ images can be seen in (c). The \( B_1^+ \) achieved its peak in the center of the sphere, reaching 108 \% of the nominal $B_1$ level, and decreased to the lowest point, 91 \%, at the edge of the sphere. The $\Delta \omega$ remained low, < 0.3 ppm, until the center, then increased to a peak of 0.12 \ ppm. Within the homogeneous $B_0$ region, the CEST contrast was comparable for OC, adia SL, and Gaussian saturation, In comparison, the BPT $MTR_{asym}$ more closely aligned with changes in \( B_1^+ \). As the $\Delta \omega$ shift increased, notable deviations emerged in the BPT. Similarly, the adia SL displayed deviations relative to the profile's low $\Delta \omega$ region. In contrast, the OC and Gaussian saturation remained relatively stable even amidst increasing frequency shifts. Notably, the OC exhibited the highest and most consistent saturation, expressing a saturation comparable to the low $\Delta B_0$ region and demonstrating a uniform saturation over the profile. \\

The Regions of Interest (ROIs) studied, each exhibiting varying degrees of $\Delta \omega$ shift, are depicted in Figure \ref{fig:ROI} (a). The corresponding statistics can be seen in the boxplots (d) and in Table \ref{MTR_statistics}. Particularly pronounced is the $\Delta \omega$ shift beneath the bubble at the sphere's apex, with additional inhomogeneities identified at the base and sides of the sphere, intersected by regions of minor shifts. In the most homogeneous ROI, ROI 1, the frequency shift was measured as $\Delta \omega =$ (0.037 $\pm$ 0.009) \ ppm. Here, the OC pulse displayed the highest exchange weighting with a peak $MTR_{asym}$ of 16.8 \%, followed by the BPT at 16.1 \%, adia SL at 15.4 \% and Gaussian saturation at 11.3 \%. The OC had the lowest Interquartile Range (IQR) at 0.6 \%. Conversely, ROI 2 represented the most inhomogeneous region with a substantial frequency shift of $\Delta \omega =$ (0.29 $\pm$ 0.04) ppm. Again, OC exhibited the highest mean $MTR_{asym}$ peak, 14.1 \%, followed by the adia SL at 13.7 \%. Notably, BPT (11 \%) and Gaussian (9.9 \%) peaks substantially declined compared to those in the center, while the IQR varied significantly between OC saturation (0.9 \%) and BPT (4 \%). ROI 3, located at the edge of the sphere, displayed an intermediate frequency shift ($\Delta \omega =$ (0.063 $\pm$ 0.034) ppm) and greater $B_1$ deviation compared to the center. The mean peak was again highest for OC saturation (15.9 \%), with BPT (15.0 \%) and adia SL (14.4 \%) following closely, and was the lowest with Gaussian saturation (10.8\ \%). ROI 4, sharing a similar $B_1$ deviation to ROI 3 but presenting the lowest $\Delta B_0$ of $\Delta \omega =$ (-0.010 $\pm$ 0.011) ppm, again OC contrast was highest (16.0 \%), with BPT (15.5 \%), SL (15.1 \%) and Gaussian (10.6 \%) in succession. In this ROI the IQR was highest for the adia SL and comparable for the other saturation pulses. \\

Figure \ref{fig:ROI} (e) illustrates the mean spectra within ROI 2. The adia SL presents asymmetric saturation around 0 ppm and a plateau between 0 and 0.5 ppm, showcased in the uncorrected pixel-wise spectra (f). The fit of the adia SL spectrum was found to have a higher discrepancy between the fitted line and the measurement points compared to the other fitted spectra in this region. Substantial deviations between 0 and 1 ppm were found within the $MTR_{asym}$ measurements of adia SL in this ROI. Similarly, the $MTR_{asym}$ of the BPT revealed considerable deviations and a shift in the peak position. The boxplot highlights high asymmetric deviations around the median for adia SL and pronounced uncertainties for BPT saturation. \\

\subsection{Magnetization vector dynamic simulation}
Figure \ref{fig:M_vs_BPT_phase_B0} (a) depicts the BPT amplitude over time, while (b) shows the OC pulse train. (c) and (d) display the Z-spectra for the two-pool simulation for a constant phase, while (e) and (f) display the spectra for an adjusted phase. The saturation of the BPT highly depends on the phase, for example, the z-magnetization at 1 ppm is 220 \% higher than the simulation with the phase adjusted after each pause. These artifacts are periodically introduced, depending on the phase mismatch at the different off-resonances $\Delta \omega$ making the spectrum non-smooth. Adjusting the phase after every pulse can avoid this, provided there are no field inhomogeneities. The OC pulse repeatedly reorients the magnetization vector in the z-direction after every block. This applies both to a constant phase and to a phase adjusted after each block, resulting in a nearly identical z-magnetization for the majority of off-resonant frequencies, which promotes a smooth spectrum (simulations of the magnetization vector over time for BPT and OC are displayed in the Supporting Information Figures S2,3). Analogously, in Figure \ref{fig:M_vs_BPT_phase_B0} (g) and (h) after adding a $B_0$ shift of 0.1 ppm to the simulations with adjusted phase after every pause, artifacts are introduced into the spectrum of the BPT while the OC spectrum remains smooth under equal conditions. \\

\section{Discussion}
The aim of this work was to optimize the pulse shape of a CEST saturation pulse train to achieve high CEST contrast and stability in the presence of field inhomogeneities. The OC pulse train was compared against CW and Gaussian saturation in phantom measurements on a 7~`T preclinical scanner and against commonly used pulse types, such as Gaussian, BPT and advanced approaches like adiabatic SL pulses in phantom measurements on a 3~T clinical scanner. Additionally, Bloch-McConnell simulations were conducted to study the magnetization vector dynamics in the water pool during saturation with OC and BPT, to understand their behavior under the influence of $B_0$ inhomogeneities. The OC (Optimal Control) framework presented in this study successfully optimized the pulse shape. This enhancement not only improved contrast but also ensured stability amidst field inhomogeneities, outperforming other pulsed saturation strategies. \\

The different presented pulsed saturation regimes show substantial impact on the resulting spectra. Since the presented pulse trains have the same $B_{1RMS}$, DC and saturation time, the only parameter that differentiates them is the time dependent $B_1(t)$ magnitude and phase. This work shows that with different pulse shapes several aspects of the CEST spectrum can be improved, suggesting the existence of an optimal pulse shape that enhances favorable features. Yoshimaru et al. were the first to successfully optimize pulse shapes, which were beyond the optimization of simple Gaussian or Sinc pulse shape parameters by applying a multi objective genetic algorithm \cite{yoshimaru2016design}. They compared the resulting pulse shapes against Gaussian and BPT / CW protocols in simulation and measurements. The primary focus of their research was on optimizing exchange weighting. Although they were successful in surpassing Gaussian, they were unable to approach saturation efficiency of CW/BPT saturation. \\

With the OC pulse designed in this work it was possible to outperform the exchange weighting of Gaussian pulse trains at 3 and 7 T in simulation and in creatine phantom measurements by up to 40-50 \%. At 7T, the proposed OC pulse was able to reach the exchange weighting of CW saturation within a few percent and exceeded at 3 T BPT exchange weighting in all measurements. \B{This performance difference also depends on the specific measurement parameters and physical environment at hand. In particular, the choice of the $B_{1RMS}$ is of crucial importance as it influences the spillover and exchange weighting, which in principle depends on parameters such as the exchange rate, the relaxation times of water and CEST pool, $B_0$ and the resonance offset. In this study creatine was used for experimental validation. With an exchange rate of several hundred Hz and an off-resonance of roughly 2 ppm, it is a representative CEST agent in phantom studies and its physical properties lying within a typical range. The choice of $B_{1RMS}$ value for our measurements was guided by the maximum $MTR_{asym}$ for CW saturation as stated by Zu et al.\cite{zu2011optimizing}, a finding corroborated by our simulations (see Figure \ref{parameter_check} (e)). Recent in vivo measurements by Wang et al. 2024 for 3 T \cite{wang2024creatine} further support this $B_{1RMS}$ range. \R{R2;C1} \\

We also conducted simulations to examine Fermi pulses, which, in terms of their shape, fall between Gaussian and block pulse trains (Supporting Information Figures S11 and S12). Compared to Gaussian saturation pulses, Fermi pulses demonstrated improved exchange weighting. However, their exchange weighting was notably less than that of CW and OC saturation. We observed that Fermi pulses introduced instabilities in the water peak within the -0.8 to 0.8 ppm range. These findings align with those reported in the study by Liu et al. \cite{liu2013nuts}. \R{R2;C2} }

\subsection{Exchange weighting and stability in the presence of field inhomogeneities}

The clinical utility of CEST effect measurements critically depends on their robustness to field inhomogeneities. Especially BPT saturation strategies can suffer from severe artifacts in the presence of $B_0$ inhomogeneities, which is especially obvious in the $B_0$ dependent signal behavior in Figure \ref{fig:3T_sphere_asym}. These artifacts could not be eliminated with $B_0$ correction and were only corrected to some extent by Lorentzian fitting. Adia SL saturation produced images with a notably higher contrast compared to Gaussian saturation, though it showed slightly less saturation than BPT. However, these images presented a subtle cloudy pattern, especially in the $B_0$ inhomogeneous region, a pattern that was further emphasized upon fitting. While adia SL saturation offers a valuable option for high-contrast and robust CEST imaging, OC saturation produced the most uniform images, coupled with the highest CEST contrast. \\

The robustness of the OC saturation to field inhomogeneities was evident in CEST creatine measurements at 3 T. In Figure \ref{fig:ROI} (c) the stability of the contrast was comparable for OC, adia SL and Gaussian saturation in the homogeneous $B_0$ region. The BPT saturation followed the \( B_1^+ \) trend more closely indicating a stronger dependence to the \( B_1^+ \). However, as the line profile surpasses the center into a increasingly $B_0$ inhomogeneous region the robustness of the OC pulse became particular evident. While the contrast of the OC pulse was similar in the $B_0$ homogeneous and inhomogeneous region, the contrast generated by the other pulses showed deviations. The Gaussian profile was comparably robust, the adia SL saturation showed deviations with increasing $B_0$ shift and the BPT expressed sever artifacts starting at even slight inhomogeneities. This also became evident by investigating the statistical deviations in different ROIs (d). Over all ROIs the OC remained stable within all combinations of field variations. The susceptibility of the adia SL to strong $B_0$ inhomogeneities as described before by \cite{herz2019cest} became evident in ROI 2. Here the statistics as well as the spectra indicated that the adia SL saturation lead to an asymmetric spectrum around the water peak. This is especially challenging as artifacts in the $MTR_{asym}$ appear. Applying a Lorentzian fit to the water peak can yield inaccurate results, as a symmetric saturation profile is assumed. While the average spectra over the BPT were smooth, and the pixelwise fit was accurate, the deviation in the $MTR_{asym}$ and the statistics became apparent. The Gaussian saturation showed comparably high but in general stable deviations in all combinations of field inhomogeneities. \\

Remarkably, the OC pulses display inherent $B_0$ robustness, a feature not explicitly included in the cost functional but naturally emerging during the process, distinguishing our approach from previous studies like \cite{graf2022advanced}. Initially, the starting pulse, a BPT with constant phase, introduced severe artifacts/wiggles (Figure 5 b and c). Yet, the optimization found a solution that was effectively independent of phase changes, thus automatically compensating for unadjusted phase or $B_0$ offsets. \\

Incorporating an MT pool in the simulation similarly reduced the $MTR_{\text{asym}}$ for OC pulses, analogous to the effects observed with Gaussian and BPT saturation.

\subsection{Insights from dynamic Bloch McConnell simulations}
Dynamic Bloch McConnell simulations reveal that the origin of artifacts introduced by BPT saturation can be traced to a mismatch in phase between the pulses. After each block the magnetization vector ends with a non-zero magnetization in the $x$- and $y$ direction. Only if the phase in the pause between the blocks is adjusted by $\omega_n T_{pulse}  B_0 \gamma$ the resulting spectrum is smooth. However, $B_0$ inhomogeneities result in a phase mismatch after each block, leading to variable levels of saturation, depending on the off resonance. This causes artifacts in the spectrum (for more detailed explanation see Supporting Information Section 2). Analogously, similar artifacts can be observed in simulations by introducing a $B_0$ inhomogeneity of 0.1 ppm. These artifacts can be minimized by reducing the pause duration. However, the DC is restricted by SAR and hardware limits. Of particular interest is the ability of the OC pulse train to effectively flip the magnetization vector in the $z$ direction, with $M_x$ and $M_y$ approaching zero after each pulse. Therefore, the subsequent pulse is independent of the previous pulse and independent of the pause duration. This makes the OC pulse inherently robust against $B_0$ inhomogeneities. \\

The reversal of the magnetization vector into the $z$ direction after a saturation period is a concept specific to SL saturation techniques. Imaging with $T_{1 \rho}$ contrast instead of conventional CEST saturation has been shown to exhibit several advantages, such as high exchange weighting and selectivity \cite{SLoverCEST,jin2011spin,yuan2012mr}. However, the off-resonant SL is susceptible to field inhomogeneities, underscoring the significance of Herz et al.'s recommendation to substitute tip-down and tip-up pulses with hyperbolic secant adiabatic pulses tailored to specific field strengths \cite{herz2019cest,herz2019t1rho}. Yet, the off-resonant SL saturation is elaborate to implement. At every off resonance $\omega_n$ the phase of the tip up and tip down is different. Furthermore, the phase is also different if the frequency shift is either $\omega_n >$ 0 ppm or $\omega_n <$ 0 ppm. This is problematic if due to $B_0$ shifts a $\omega_n >$ 0 ppm SL pulse is played at a $\omega_n <$ 0 ppm point and vise versa. This makes the pulse vulnerable in the region around $\omega =$ 0 ppm, which leads to the asymmetric saturation as demonstrated in Figure \ref{fig:ROI}. It appears that the other investigated pulses do not share this problem. Furthermore, to ensure robust tipping, the adiabatic pulses need to be long enough and they are generally SAR intensive. This limits the time and energy available for CEST saturation. \\

Interestingly, magnetization vector dynamics simulations suggest that OC pulses employ a saturation strategy similar to the off-resonant SL technique (this can be seen in the Supporting Information Figure S3). This process involves a tipping down phase, a saturation/locking phase, and a tipping back to $M_x, \  M_y = 0$ (for visualization see the videos of OC pulse saturation dynamic in the Supporting Information Video S1,2). Notably, OC pulses also achieve saturation during the tipping phase, thereby optimizing energy and time utilization.

\subsection{Considerations for OC optimization}
\begin{figure} [H]
    \begin{center} 
        \includegraphics[width=1\textwidth, trim={7cm 1cm 6cm 2cm}]{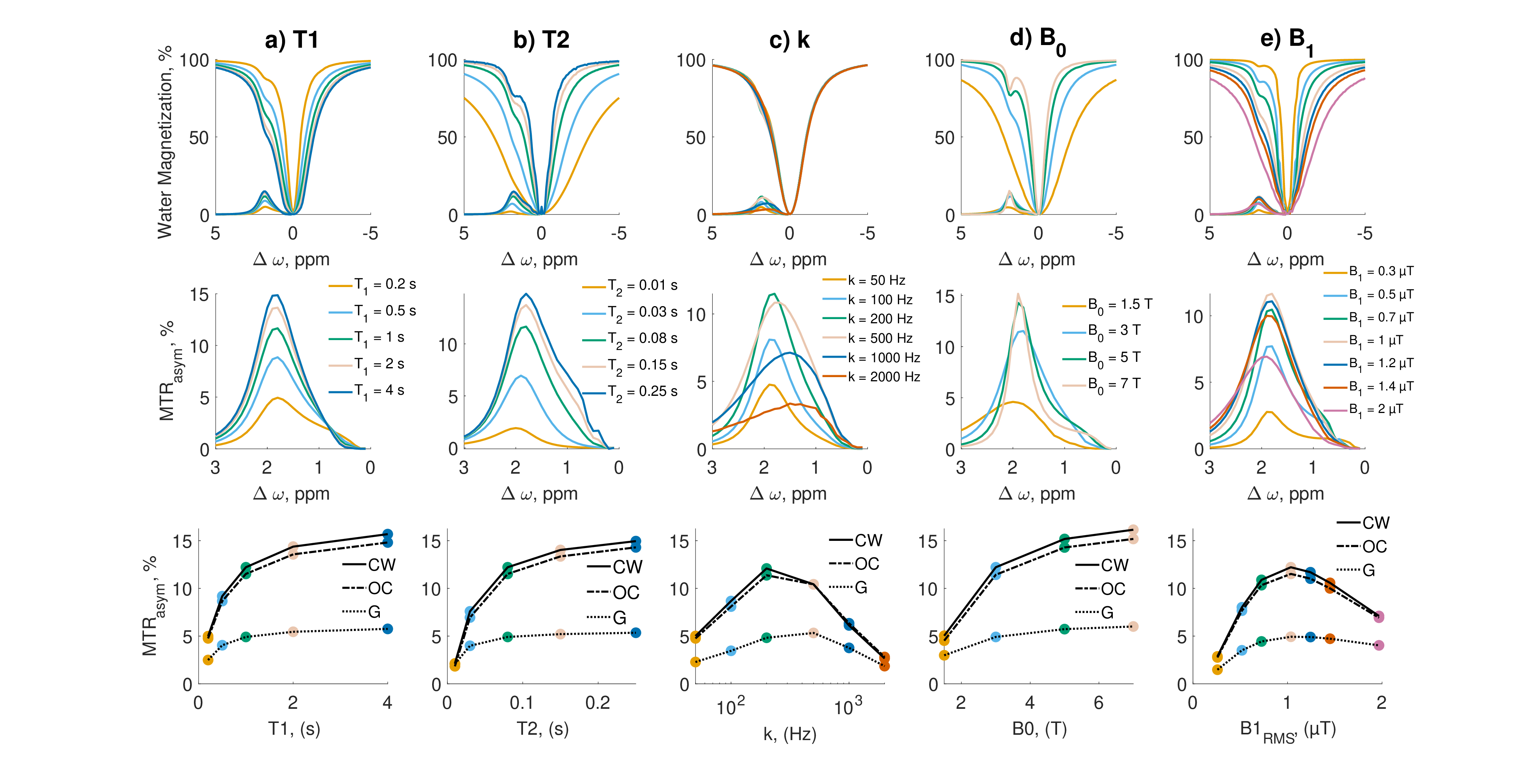}
    \end{center}
        \caption{Simulation of OC Saturation for varying (a) $T_1$ times, (b) $T_2$ times, (c) exchange rates $k$, (d) field strengths $B_0$, and (e) OC pulse amplitude scaling. The first row displays simulated spectra for OC saturation, while the second row shows the corresponding $MTR_{asym}$. The third row presents the peak $MTR_{asym}$ values, comparing OC with CW and Gaussian simulations under the same physical conditions and pulse parameters.}
    \label{parameter_check}
\end{figure}

\begin{figure} [H]
    \begin{center} 
        \includegraphics[width=1\textwidth, trim={1cm 0cm 0cm 0cm}]{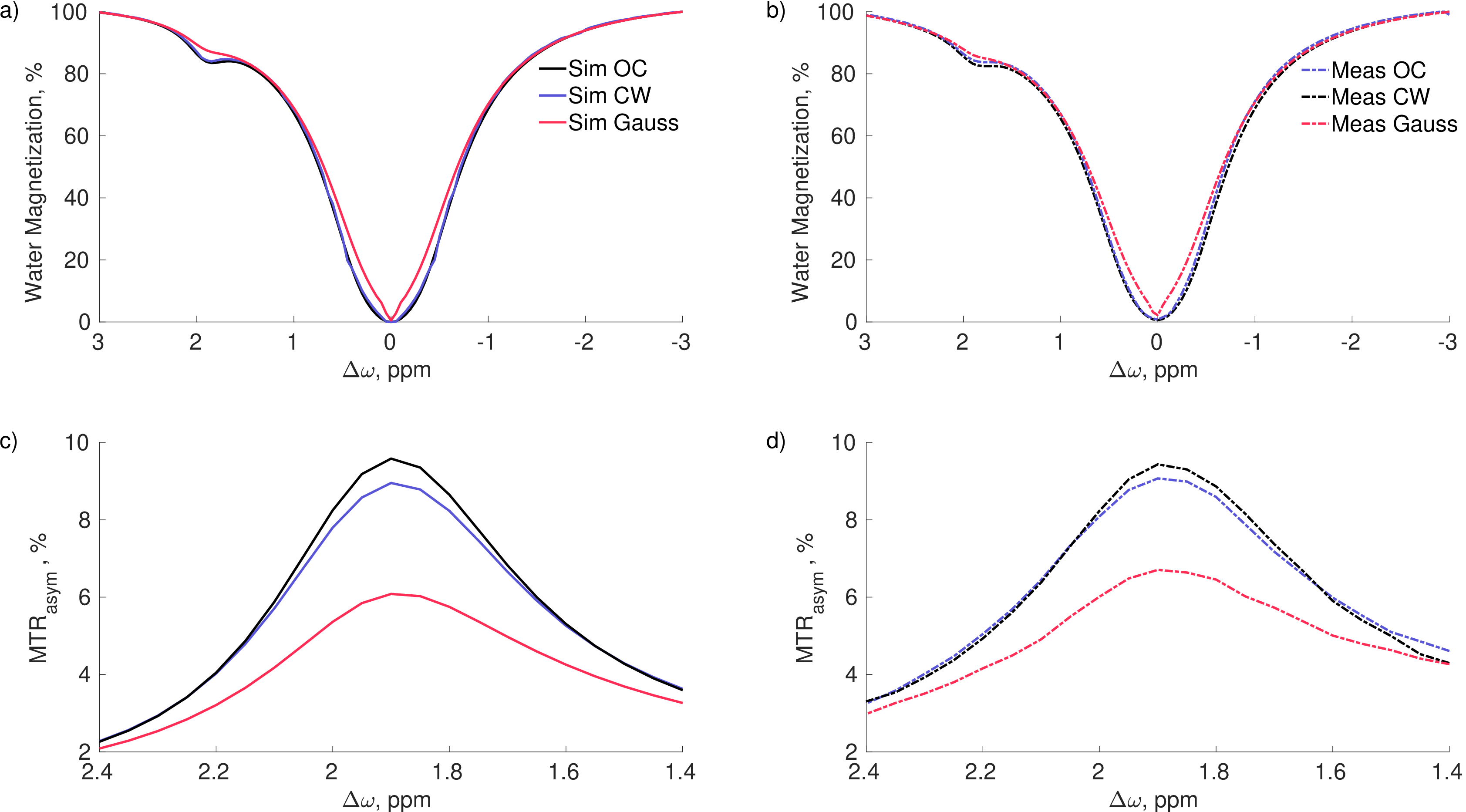}
    \end{center}
        \caption{Comparison of simulation and measurements in a creatine phantom at 7T. Simulated spectrum for CW, Gauss and OC saturation (a) with simulated $MTR_{asym}$ peak in (c). Uncorrected measurement spectrum for the same pulses (b) with corresponding $MTR_{asym}$ in (d).}
    \label{fig:7T}
\end{figure}

\begin{figure} [H]
    \begin{center} 
        \includegraphics[width=1\textwidth, trim={10cm 1cm 5cm 1cm}]{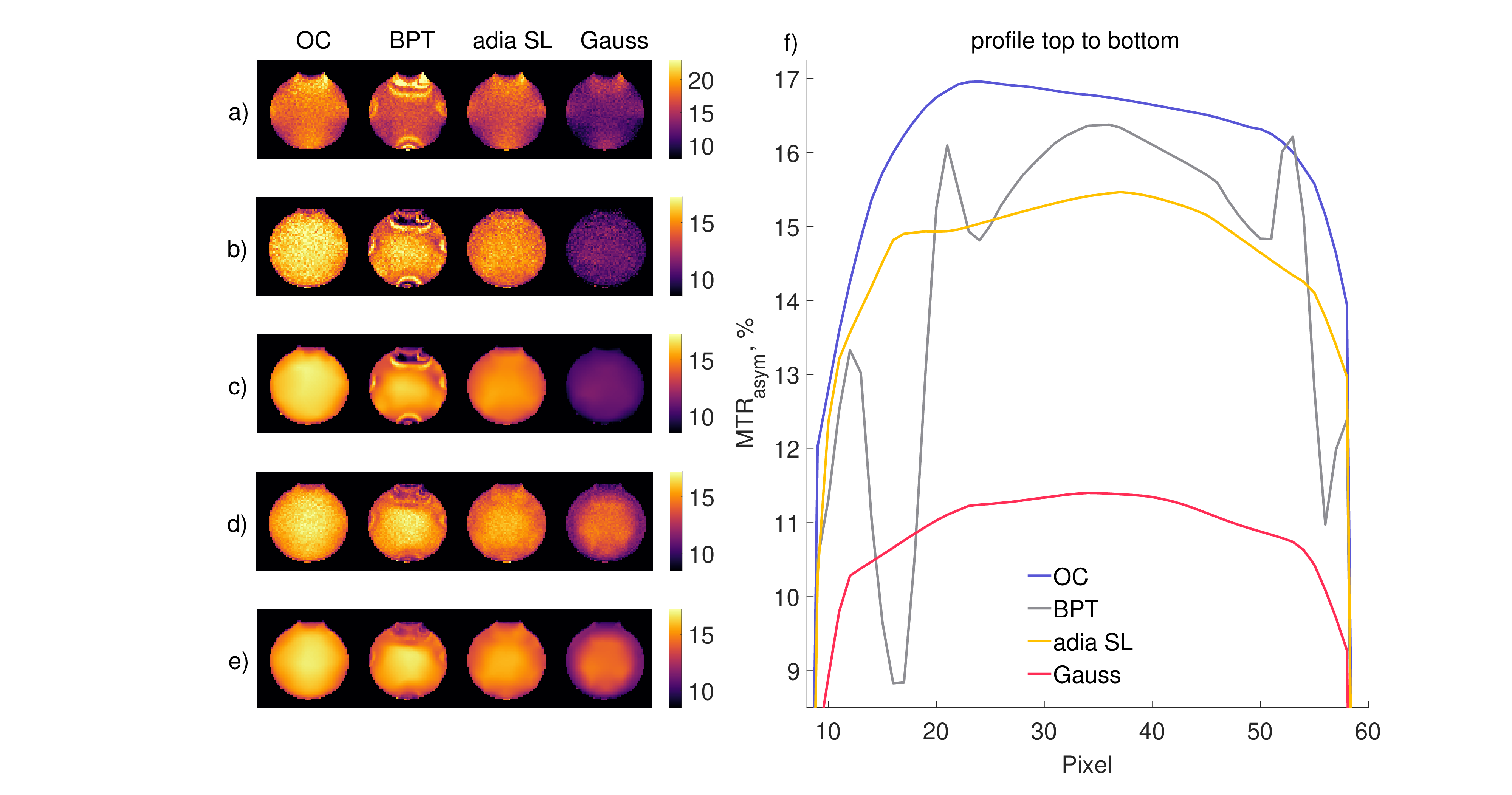}
    \end{center}
        \caption{For each pulse the (a) uncorrected $MTR_{asym}$, (b) $B_0$-corrected $MTR_{asym}$, (c) TGV-denoised and $B_0$-corrected $MTR_{asym}$, (d) amplitude of the creatine CEST-peak in the two-pool-Lorentzian-fit and (e) TGV-denoised two-pool-Lorentzian-fit images is shown. An air bubble is visible at the top of the images introducing $B_0$-inhomogeneities. (f) Vertical creatine CEST signal profile from top to bottom through the $B_0$-inhomogeneities in (c). }
    \label{fig:3T_sphere_asym}
\end{figure}

\begin{figure} [H] 
        \begin{center}
        \includegraphics[width=1\textwidth, trim={1.5cm 2.0cm 1.35cm 5cm}]{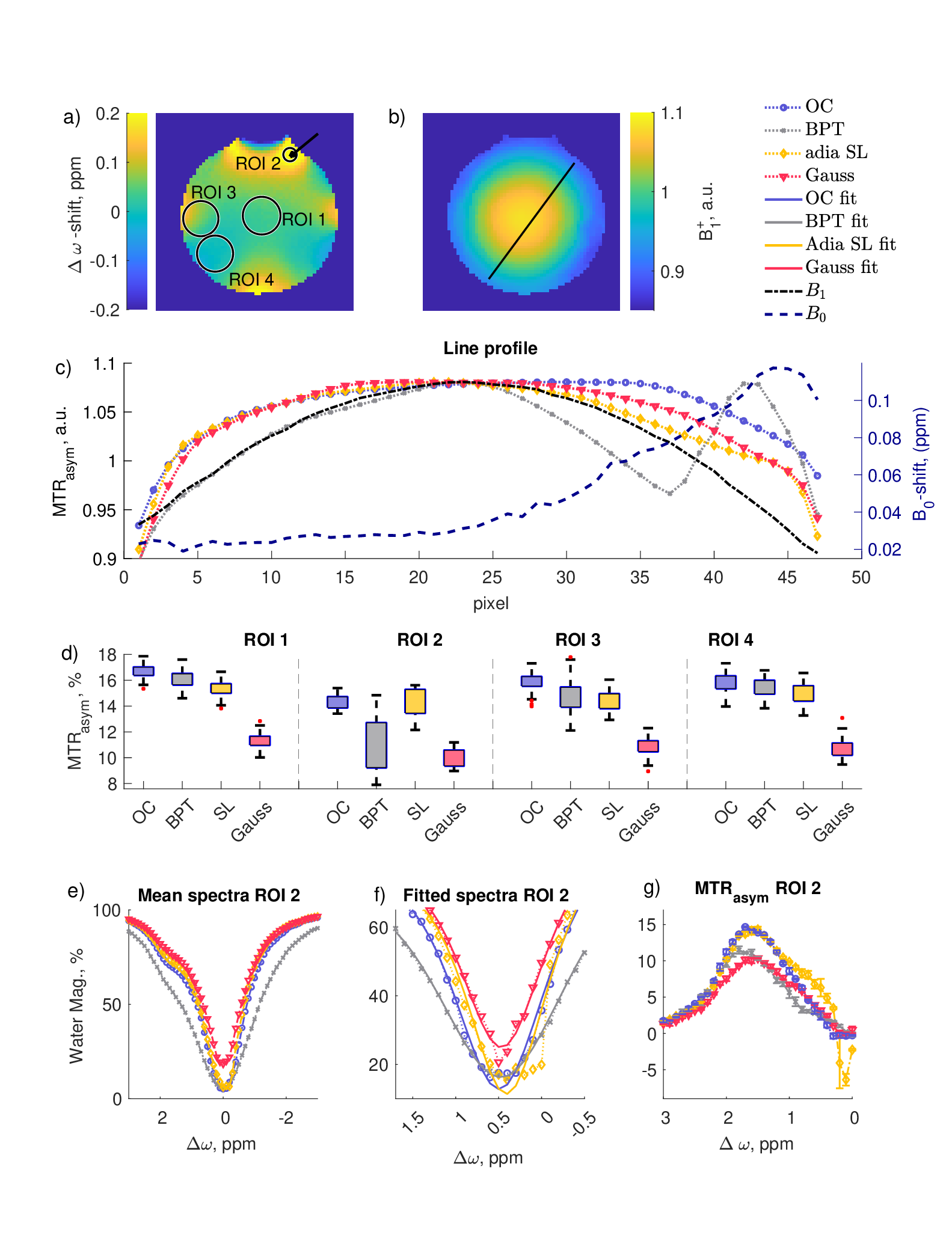}
        \end{center}
        \caption{OC, BPT, adia SL and Gauss creatine phantom CEST spectra measured at 3 T . (a) $\Delta \omega$ shift map due to $B_0$ inhomogeneities with ROIs with different levels of field inhomogeneities. (b) Bloch Siegert $B_1$-map divided by the nominal $B_1$ level. (c) Line profile in (b) through the center of the $B_1$ map and the most homogeneous $B_0$ region. (d) Boxplots of the the different ROIs in (a). (e) Mean spectrum in ROI 2. (f) Spectrum and fit of a pixel in ROI 2 (see arrow). Gauss) $MTR_{asym}$ of the spectra in (e).}
        \label{fig:ROI}
\end{figure}

The simulations in Figure \ref{parameter_check} indicate that the OC framework effectively generates pulse shapes for diverse parameter sets, producing results with exchange weighting and stability comparable to CW saturation. The pulse operates stable with changes in $T_1$. The simulations indicate that the pulses can be used independently of the scanner field strength, which was demonstrated at a 3 T clinical system, and a 7 T preclinical system. This suggests that the optimized pulses are suitable for low to ultra high field strengths. 

While the optimized pulses produce satisfying results for a broad range of parameters, they are not without limitations. Simulations indicate that high $T_2$ (> 250 ms) values and strong RF amplitude scaling (> 200 \%) might introduce artifacts in the spectrum, although the $MTR_{asym}$ peak value remains stable. For applications with higher $B_1$ saturation, optimization of a pulse for a new target spectrum might be necessary. The wiggles at $k = 2000 s^{-1}$ in Figure \ref{parameter_check} (c) can be removed by optimizing for a higher exchange rate (for details see Supporting Information Figure S4).  \\

The pulses were optimized for a DC of 90 \% with single pulse duration of 100 ms. As can be seen in simulation Figure \ref{fig:M_vs_BPT_phase_B0} the pulse is robust against phase variations occurring during the pauses. This implies that the DC, and hence pauses, can be freely adjusted. However, a specific optimization is needed if a different pulse length $t_d$ is desired, as the single saturation duration is fixed here. Chaining OC pulse trains can be used to increase total saturation time. Specifically, this allows for the total saturation time to be extended to integer multiples of the original $T_{sat}$ (for an example see Supporting Information Figure S10). \\

Optimization quality was improved by adopting real-valued pulses, which, unlike complex pulses, produce a symmetric spectrum, thereby reducing spectral artifacts. When given the freedom to optimize the phase, the optimizer found pulses that artificially enhanced the CEST effect. This means that we were able to measure a CEST peak in pure water even without a CEST agent. We are aware that this decision reduced the versatility of possible outcomes. Allowing phase optimization along the pulse train could potentially enhance desirable CEST saturation attributes, such as improved \( B_1^+ \) robustness. Incorporating the phase into the optimization is a logical next step for further research. \\

The correct choice of the number of $\Delta \omega$ in the z-spectrum was essential for the optimization. Although the optimizer produced solutions fulfilling all requirements at the prescribed points, oscillations between the optimized $\Delta \omega$ points could be observed with higher sampling rate. $\Delta B_0$ inhomogeneities displaced measurement points to areas with high oscillations, causing artifacts in the phantom spectra. However, these artifacts were minimized by employing a spectral resolution of at least 50 $\Delta \omega$ points per ppm during optimization. \\

\section{Conclusion}
We have demonstrated that with the presented optimal control optimization RF pulse design method it is possible to design the pulse shape of an RF pulse for CEST imaging. On a 7 T preclinical scanner system, the optimized OC pulse train achieved saturation comparable to CW saturation and outperformed Gaussian saturation by up to 50 \% in phantom measurements. The OC pulse train surpassed state-of-the-art CEST saturation strategies, producing the highest CEST contrast and demonstrating highest stability against field inhomogeneities in both simulation and 3 T creatine phantom measurements.

\section{Conflict of Interest}
The authors declare no competing interests.
\section{Open Research}
Optimized pulses and scripts for reconstructing all result figures will be accessible after publication at: https://gitlab.com/CleStil/enhanced-and-robust-contrast-in-cest-mri
\section{Acknowledgment}
This research was funded in whole, or in part, by the Austrian Science Fund (FWF) I4870. For the purpose of open access, the author has applied a CC BY public copyright license to any Author Accepted Manuscript version arising from this submission.

\section{Caption Video}
Video S1: Evolution of the water magnetization vector in a two-pool Bloch-McConnell simulation using the proposed OC pulse at an off resonance of $\Delta \omega$ = 1 ppm. The simulation demonstrates how the magnetization vector is rapidly aligned to achieve $M_x = 0$ and $M_y = 0$ after each pulse, enhancing robustness against B0 field inhomogeneities. Simulation based on \ref{fig:M_vs_BPT_phase_B0} (d).\\
Video S2: This simulation extends Video S1’s scenario, showcasing the OC pulse with phase adjustments after each pulse, similar to the BPT method. Simulation based on \ref{fig:M_vs_BPT_phase_B0} (h).

\listoffigures

\pagebreak

\definecolor{mygray}{gray}{0.6}
\begin{table}[H] 
    \centering 
    \caption{Results from Figure \ref{fig:ROI} d, statistics for the median and IQR of the \(MTR_{\text{asym}}\) in the ROIs 1-4 with different levels of \(B_0\) inhomogeneities.}
    \begin{tabular}{lll|cc}

        & & & \(MTR_{\text{asym}}\), \% & \(IQR\), \% \\ 
        \midrule
        \multirow{4}{*}{ROI 1} & \multirow{4}{*}{$\Delta \omega =$ (0.037 $\pm$ 0.009) ppm} & OC & 16.8 & 0.6 \\
        & & BPT & 16.1 & 0.9 \\
        & & adia SL & 15.4 & 0.7 \\
        & & Gauss & 11.3 & 0.7 \\
        \midrule
        \multirow{4}{*}{ROI 2} & \multirow{4}{*}{$\Delta \omega =$ (0.29 $\pm$ 0.04) \ ppm} & OC & 14.1 & 0.9 \\
        & & BPT & 11 & 4 \\
        & & adia SL & 13.7 & 1.9 \\
        & & Gauss & 9.9 & 1.2 \\
        \midrule
        \multirow{4}{*}{ROI 3} & \multirow{4}{*}{$\Delta \omega =$ (0.063 $\pm$ 0.034) \ ppm} & OC & 15.9 & 0.7 \\
        & & BPT & 15.0 & 1.5 \\
        & & adia SL & 14.4 & 1.1 \\
        & & Gauss & 10.8 & 0.8 \\
        \midrule
        \multirow{4}{*}{ROI 4} & \multirow{4}{*}{$\Delta \omega =$ (-0.010 $\pm$ 0.011) \ ppm} & OC & 16.0 & 1.0 \\
        & & BPT & 15.5 & 1.0 \\
        & & adia SL & 15.1 & 1.2 \\
        & & Gauss & 10.6 & 0.9 \\
        \bottomrule
    \end{tabular}
    \label{MTR_statistics}
\end{table}

\pagebreak

\B{
\begin{figure} [H]
    \begin{center} 
        \includegraphics[width=1\textwidth, trim={1cm 2cm 1cm 1cm}]{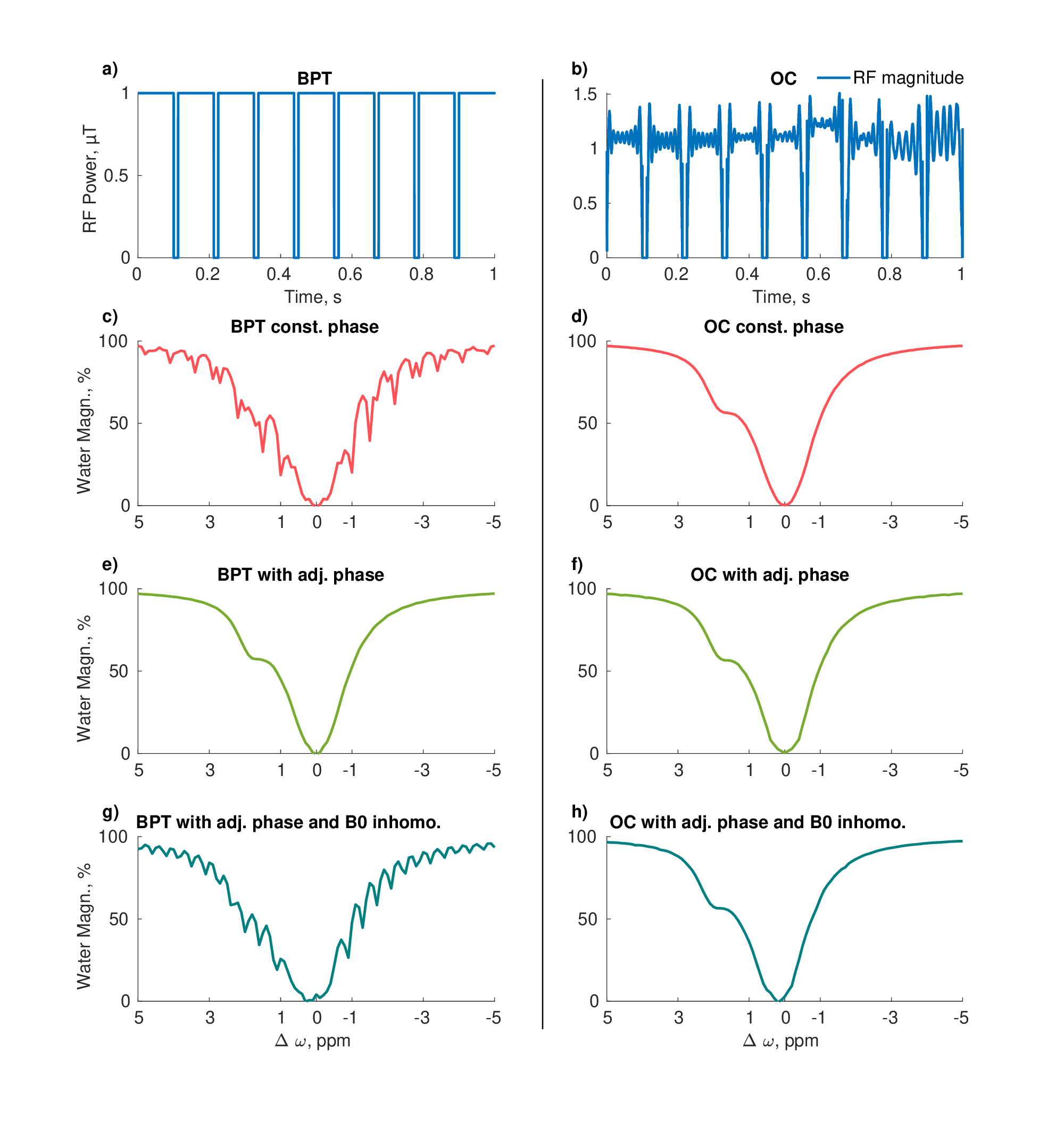}
    \end{center}
        \caption{Comparison of BPT (a) and the OC pulse train (b) simulations under different conditions: with constant phase (c,d), with an adjusted phase by $\omega_n T_{pulse} B_0 \gamma$ after every pause (e,f) and with an adjusted phase and an additional $B_0$ inhomogeneity of 0.1 ppm.}
    \label{fig:M_vs_BPT_phase_B0}
\end{figure}
}


\end{document}